\documentclass[two-column, nofootinbib]{revtex4-1}

\usepackage{amsmath}
\usepackage{tabularx}
\usepackage{graphicx}
\usepackage[usenames,dvipsnames,svgnames]{xcolor}
\usepackage{hyperref}

\hypersetup{dvips,dvipdfm,colorlinks=true,urlcolor=magenta,filecolor=magenta,linktoc=page,citecolor=red,linkcolor=blue,bookmarks=true}

\begin{document}

\title{Viaggiu entropy and the generalized second law in a flat FLRW Universe}

\author{Subhajit Saha\footnote {Electronic Address: \texttt{\color{blue} subhajit1729@gmail.com}}}
\affiliation{Department of Mathematics, \\ Panihati Mahavidyalaya, Kolkata 700110, West Bengal, India.}


\begin{abstract}

\begin{center}
(Dated: The $13^{\text{st}}$ November, $2019$)
\end{center}

Our aim, in this paper, is to study the generalized second law by considering the dynamical apparent horizon to be endowed with the Hawking temperature and the Viaggiu entropy introduced in a pioneering work in 2014. We have devoted our attention to a flat FLRW universe filled with a perfect fluid having a constant equation of state $p=w\rho$. It is worthwhile to note that we have considered both forms of the Hawking temperature, the original one as well as the truncated version. The latter one is generally used during calculations, however, several arguments have been put forward against it. Our analysis yields a startling result. The Viaggiu entropy naturally forbids the phantom era. In other words, the Viaggiu entropy will never allow the EoS parameter to go beyond the Cosmological Constant. This result is in strong agreement with recent observations and we have deduced it purely by thermodynamic means. This is in striking contrast with the results obtained with the Bekenstein-Hawking formalism.\\\\ 
Keywords: Dynamical apparent horizon; Viaggiu entropy; Hawking temperature; Generalized second law\\\\
PACS Numbers: 98.80.-k\\\\

\end{abstract}

\maketitle


\section{Introduction}
The study of horizons in the context of Cosmology has gained much attention in recent years thanks to the establishment of the intimate connection among gravitation, thermodynamics, and quantum theory resulting from four decades of extensive research in the theory of black holes (BHs). This relationship has helped shape the laws of BH thermodynamics which show that BHs obey certain laws which are analogous to the ordinary laws of thermodynamics at purely classical level. For an excellent review on the latter topic, one may see the article by Wald \cite{Wald1} and the book by Poisson \cite{Poisson1}. It is interesting to note that such a beautiful analogy has been made possible due to two pioneering hypotheses put forward by J. Bekenstein and S. Hawking respectively. In 1973, Bekenstein \cite{Bekenstein1} proposed that the entropy associated with the event horizon of a BH is proportional to the area of the BH bounded by it. Two years later, Hawking \cite{Hawking1} found that BHs are not literally "black" rather they are capable of radiating like a black body due to quantum effects of particle creation. Moreover, this temperature is proportional to the surface gravity on the event horizon of the BH. Thus, the entropy $S_h$, known as the Bekenstein entropy, and the temperature $T_h$, known as the Hawking temperature, associated with a BH horizon $R_h$ is mathematically represented as
\begin{equation} \label{be}
S_h=\left(\frac{c^3}{G\hbar}\right)\frac{A_h}{4},
\end{equation}
and
\begin{equation} \label{ht}
T_h=\left(\frac{\hbar c}{k_B}\right)\frac{|\kappa _h|}{2\pi}
\end{equation}
respectively, where $c$ is the velocity of light, $\hbar$ is the reduced Planck's constant, and $k_B$ is the Boltzmann's constant. $\kappa _h$ and $A_h$ are the surface gravity on the horizon and the area bounded by the horizon respectively. The Hawking temperature and the Bekenstein entropy are related to the mass of the BH through the first law of BH thermodynamics. As the temperature and the entropy are characterized by the spacetime geometry and hence by the Einstein field equations, so it is natural to speculate that there is some connection between BH thermodynamics and Einstein equations. Indeed, it is worth noting that Jacobson \cite{Jacobson1} used the first law of BH thermodynamics to derive the Einstein's field equations (EFE) for assuming local Rindler horizons endowed with Unruh temperature measured by an accelerated observer located just inside the horizon. Subsequently, for a general static spherically symmetric spacetime, Padmanabhan \cite{Padmanabhan1} derived the first law of BH thermodynamics from the EFE.\\ 

Since the early proposals of 't Hooft \cite{Hooft1} and Susskind \cite{Susskind1}, attempts have been made to incorporate this remarkable equivalence between the dynamics of our observable Universe and that of BHs. Indeed, spacetimes admitting horizons exhibit a resemblance to thermodynamic systems which allows us to associate the notions of temperature and entropy with them, similar to the case of BHs. Bak and Ray \cite{Bak1}, in 1999, were the first to extend this formalism to Cosmology and who stressed the role of the apparent horizon in this context. The analogy between the thermodynamic laws and the gravitational thermodynamics of horizons is quite intriguing, although it is not understood at a deeper level yet \cite{Padmanabhan2,Padmanabhan3}. However, Paranjape, Sarkar, and Padmanabhan \cite{Paranjape1} proposed that these results can be interpreted by hypothesizing that spacetime corresponds to an elastic solid and that its equations of motion are analogous to those of elasticity. Now, since the laws of thermodynamics are considered to be universally true, it is generally believed that those cosmological models which are consistent with the laws of thermodynamics in addition to being consistent with observational tests are more acceptable as compared to those models which only pass the observational tests. Although there exist several horizons in Cosmology which may be relevant to the study of thermodynamics, we shall confine ourselves to apparent horizons only mostly because their existence is universal and partly because they bring a high degree of mathematical simplicity into calculations when we consider a flat Friedmann-Lemaitre-Robertson-Walker (FLRW) universe. It is worth mentioning that the equivalence of the first law of thermodynamics (FLT) and the Friedmann equation on the apparent horizon in Einstein gravity as well as in the Gauss-Bonnet and Lovelock gravity theories was first put forward by Cai and Kim \cite{Cai1}. Then, Akbar and Cai \cite{Akbar1} extended the study to scalar-tensor gravity and $f(R)$ gravity theories. The FLT was obtained from the Friedmann equation on the apparent horizon in different theories of gravity, including in the Einstein, Lovelock, nonlinear, and scalar-tensor theories by Gong and Wang \cite{Gong1} with the help of a mass-like function having the dimensions of energy and equal to the Misner-Sharp-Hernandez (MSH) mass on the apparent horizon. Interested readers may see the book by V. Faroni \cite{Faraoni1} for a detailed account of the features of gravitational thermodynamics.\\

In this paper, we wish to study the generalized second law (GSL) in a flat FLRW universe filled with a perfect fluid having a constant equation of state $p=w\rho$. We shall concentrate on the dynamical apparent horizon endowed with the Hawking temperature and the Viaggiu entropy. It is worthwhile to mention that the latter was introduced in 2014 by S. Viaggiu and bears several intriguing features. Moreover, we have considered both forms of the Hawking temperature, the original one as well as the truncated version. In Sect. II, we devote our attention to the Bekenstein-Hawking formalism at the dynamical apparent horizon. Sect. III introduces the Viaggiu formalism. In Sect. IV, we have analyzed the GSL with the Viaggiu entropy and both forms of the Hawking temperature separately. Finally, Sect. V features a short discussion of the obtained results.

\section{The Bekenstein-Hawking Formalism}
We shall now quickly go through the basic ideas required to undertake a thermodynamic study on the dynamical apparent horizon. It is worthwhile to mention here that this kind of thermodynamic study of dynamical horizons was first introduced by Wang et al. \cite{Wang1} and extended later by Saha and Chakraborty \cite{Saha1,Saha2,Chakraborty1}. Consider a flat FLRW universe which is based on the assumption of the Cosmological Principle and governed by the metric
\begin{equation} \label{flrw}
ds^2=-c^2dt^2+a^2(t)\left[dr^2+r^2(d{\theta}^2+\mbox{sin}^2\theta d{\phi}^2)\right],
\end{equation}   
where $a(t)$ is the scale factor of the Universe. This metric can be locally expressed in the form
\begin{equation}
ds^{2}=h_{i j} (x^{i}) dx^{i} dx^{j}+R^{2} d\Omega_2^{2},
\end{equation}
where $i$, $j$ can take values $0$ and $1$, and $d\Omega_2^{2}=d{\theta}^2+\mbox{sin}^2\theta d{\phi}^2$, while $R=ar$ is interpreted as the proper radius. The two-dimensional metric
\begin{equation}
d\gamma^{2}=h_{i j}(x^{i})dx^{i} dx^{j},
\end{equation}
where
\begin{equation}
h_{i j}=\mbox{diag}\left(-1, a^2\right)
\end{equation}
is known as the normal metric. If we construct another scalar associated with this normal space as
\begin{eqnarray} \label{chix}
\chi &=& h^{i j}(x^{i})\partial_{i}R \partial_{j}R \nonumber \\
&=& 1-H^2R^2,
\end{eqnarray}
then the dynamical (cosmological) apparent horizon ($R_A$) of a comoving observer is defined, in mathematical terms, as a spherical surface located at the vanishing of this scalar, i.e., $\chi =0$, and we obtain
\begin{equation} \label{rapp}
R_A=\frac{1}{H}.
\end{equation}
Geometrically, an apparent horizon is the surface on which the congruence of inward null geodesics vanish so that the events outside the horizon are inaccessible to the comoving observer at $r=0$. This is in contrast to the BH event horizon which is defined as the surface where the congruence of outward null geodesics vanish so that the observer located outside the BH is unable to have causal connection with events happening inside the BH. Now, the surface gravity on the cosmological apparent horizon, $\kappa _A$, is given by
\begin{equation}
\kappa_{A}=-\frac{1}{2}\frac{\partial \chi}{\partial R}\bigg\vert _{R=R_{A}}.
\end{equation}
This definition is based on the Kodama-Hayward formulation \cite{Kodama1,Hayward1} of surface gravity for dynamical BHs. The Bekenstein entropy and the Hawking temperature on the apparent horizon are then written as\footnote{Without any loss of generality, we assume that the universal constants $c$, $\hbar$, $G$, $\kappa_B$ are all unity.}
\begin{eqnarray} \label{beapp}
S_A &=& \left(\frac{c^3}{G\hbar}\right)\frac{A_A}{4} \nonumber \\
&=& \pi R_{A}^{2}
\end{eqnarray}
and
\begin{eqnarray} \label{htapp-1}
T_A &=& \left(\frac{\hbar c}{k_B}\right)\frac{|\kappa _A|}{2\pi} \nonumber \\
&=& \frac{1}{2\pi R_A}\left(1-\frac{\dot{R}_A}{2HR_A}\right)
\end{eqnarray}
respectively. Note that Eq. (\ref{htapp-1}) reduces to 
\begin{equation} \label{htapp-2}
T_A = \frac{1}{2\pi R_A}\left(1-\frac{\dot{R}_A}{2}\right)
\end{equation}
in a flat FLRW universe. At this point, it must be mentioned that since the amount of energy crossing the apparent horizon needs to be evaluated in an infinitesimal time interval, the horizon is assumed to be slowly evolving over that time interval so that $\dot{R}_A$ may be assumed to be zero\footnote{Several arguments have been put forward \cite{Binetruy1,Helou1} against using this type of formalism though.}. In that case, Eq. (\ref{htapp-2}) reduces to
\begin{equation} \label{htapp-3}
T_A = \frac{1}{2\pi R_A}.
\end{equation}
Henceforth, in this paper, we shall call Eq. (\ref{htapp-2}) as the Hawking temperature and Eq. (\ref{htapp-3}) as the truncated Hawking temperature.

\section{The Viaggiu Formalism}
In 2014, S. Viaggiu \cite{Viaggiu1} obtained a generalized expression for the Bekenstein entropy by using suitable theorems for BH formation in Friedmann expanding universes. The modified expression contains an extra term proportional to the Hubble parameter $H$ and this term is a contribution due to the dynamical degrees of freedom of the expanding universe. The most intriguing fact about this new term is that it arises without any reference to the analogy between EFE and the laws of thermodynamics. This new proposal helps to write down the equation of state (EoS) of a BH (see Eq. (15) in Ref. \cite{Viaggiu1}). The trick is to start with the condition for the non-formation of trapped surfaces which is given by \cite{Brauer1,Koc1,Brauer2,Malec1}
\begin{equation} \label{ineq-1}
\delta M \frac{G}{c^2} < \frac{L}{2} + \frac{AH}{4\pi c}
\end{equation}  
for a sphere $S$ with proper mass excess $\delta M$. $L$ and $A$ are the proper radius and the area of the sphere respectively, $H$ is the Hubble parameter, $G$ is the universal gravitational constant, and $c$ is the velocity of light. Inequality (\ref{ineq-1}) is the starting point of out analysis. By using the well-known Bekenstein entropy bound in the spherical case, $S \leq S_{\text{max}} = \frac{2\pi \kappa_B RE}{\hbar c}$, with inequality (\ref{ineq-1}) and assuming $H=0$, one obtains the entropy for the flat asymptotically flat spacetimes. Here, $L$ is assumed to be the proper length of event horizon the BH. For $H \neq 0$ (dynamical horizon), we employ the same Bekenstein bound of the static asymptotically flat case, but on the bound on the mass-energy given by Eq. (\ref{ineq-1}). However, it must be noted that, as suggested by Hayward \cite{Hayward2,Hayward3}, BHs in expanding universes are characterized by the existence of apparent horizons. Hence, in this case, $L$ is identified as the outer apparent horizon of the BH. Then, one obtains
\begin{eqnarray} \label{entropy-2}
S_{BH} &=& \left(\frac{1}{4L_{p}^{2}}\right) A_h + \left(\frac{3\kappa _B}{2cL_{p}^{2}}\right) V_h H \nonumber \\
&=& \frac{A_h}{4} + \frac{3 V_h H}{2},
\end{eqnarray}
where $L_{p} = \sqrt{\frac{G\hbar}{c^3}}$ is the Planck length. Physically speaking, the term proportional to $V_h H$ in Eq. (\ref{entropy-2}) is a consequence of the degrees of freedom related to the dynamical nature of the spacetime. In a subsequent paper \cite{Viaggiu2}, Viaggiu extended this pioneering result to Cosmology and obtained a generalized form for the entropy at the dynamical apparent horizon of a flat FLRW universe. Furthermore, he was able to obtain a generalized expression for the internal energy $U$ and when this expression is applied to the apparent horizon of the Universe, the internal energy $U$ is found to be a constant of motion, i.e., $dU=0$\footnote{Note that this internal energy $U$ is referred to the internal energy of the fluid evaluated at the apparent horizon and not the internal energy of the apparent horizon itself.}. This result is in strong agreement with the holographic principle.

\section{The Present Work}
The aim of the present work is to study the thermodynamic implications of the corrected form of the Bekenstein entropy, as proposed by Viaggiu, on the dynamical apparent horizon of a flat FLRW universe. In particular, we are interested in studying the generalized second law (GSL) in a flat FLRW universe filled with a perfect fluid having a constant EoS $p=w\rho$, where $w>-\frac{1}{3}$ and $w \leq -\frac{1}{3}$ correspond to pre-dark energy and dark energy (DE) eras respectively. The DE era is further subdivided into quintessence era, cosmological constant, and phantom era according as $-1<w<-\frac{1}{3}$, $w=-1$, and $w<-1$ respectively. In order to realize our purpose, we need to evaluate the fluid entropy for a Universe bounded by the dynamical apparent horizon. For this, we employ the Clausius relation
\begin{equation}
T_{fA}dS_{fA} = dU + pdV_A,
\end{equation}
where $T_{fA}$ and $S_{fA}$ are, respectively, the temperature and the entropy of the fluid, $U = \frac{4}{3}\pi R_{A}^3 \rho$ is the internal energy of the fluid, evaluated at the dynamical apparent horizon, and $V_A = \frac{4}{3}\pi R_{A}^{3}$ is the volume of the fluid bounded by the dynamical apparent horizon. In accordance with the striking result obtained by Mimoso and Pavon in Ref. \cite{Mimoso1}, we may safely assume that the temperature of the horizon and that of the fluid are equal\footnote{If we do not consider them to be equal, then a temperature gradient will be generated and non-equilibrium thermodynamics will come into play.}. If we consider the Bekenstein entropy given in Eq. (\ref{beapp}) and the truncated Hawking temperature given in Eq. (\ref{htapp-3}) and note that $\dot{R}_A = \frac{3}{2}\left(1+w\right)$, then the first order time-derivative of the total entropy $S_{TA}$ is evaluated as \cite{Saha1}
\begin{eqnarray} \label{ds-b-1}
\dot{S}_{TA} &=& \dot{S}_A + \dot{S}_{fA} \nonumber \\
&=& \frac{9\pi}{2} \left(1+\frac{p}{\rho}\right)^2,
\end{eqnarray}
which reduces to
\begin{equation} \label{ds-b-2}
\dot{S}_{TA} = \frac{9\pi R_A}{2} (1+w)^2
\end{equation}
in a flat FLRW universe with a constant EoS. Eq. (\ref{ds-b-2}) shows that the GSL holds without any restriction on $w$ since $\dot{S}_{TA} \geq 0$ for all $w$. It must be noted that the expression of $T_A$ in Eq. (\ref{htapp-3}) and that of $\dot{R}_A$ can be used to obtain Eq. (\ref{ds-b-2}) without ambiguity. Now, on the other hand, if we consider the Bekenstein entropy given in Eq. (\ref{beapp}) and the Hawking temperature given in Eq. (\ref{htapp-2}), then
\begin{equation}
\dot{S}_{TA} = 18\pi R_A \frac{(1+w)^2}{(1-3w)},
\end{equation}
which implies that the GSL holds for $w \leq \frac{1}{3}$, that is, GSL is satisfied in every epoch except during the pre-radiation era. This is a new result and is relevant in the present context since we wish to compare our results obtained with the Viaggiu entropy with those obtained with the Bekenstein entropy for both versions of the Hawking temperatures given in Eqs. (\ref{htapp-2}) and (\ref{htapp-3}). 

If we now consider the Viaggiu entropy given in Eq. (\ref{entropy-2}) and the truncated Hawking temperature given in Eq. (\ref{htapp-3}) and note that $dU=0$, then
\begin{equation}
\dot{S}_{TA} = \frac{3\pi R_A}{2}(1+w)(8+3w),
\end{equation}
from which it is evident that the GSL holds either if $w \geq -1$ or if $w \leq -\frac{8}{3}$. We have, thus, arrived at a startling result that phantom era is forbidden in this scenario, since, whenever the phantom EoS ($w < -1$) is encountered, the GSL breaks down. Since GSL is an energy conservation law, any physical system, including the Universe should respect this law. {\it Thus, the Viaggiu entropy, which is a correction to the Bekenstein entropy due to the dynamic nature of the Universe, naturally forbids the phantom era.} This suggests that, if our Universe is filled with a perfect fluid with a constant EoS parameter $w$, then $w$ can never go beyond the Cosmological Constant. Recent observations also seem to agree with this theoretical result, which we have arrived at by purely thermodynamic means. In fact, a phantom fluid is plagued by series drawbacks. Such a fluid violates the dominant energy condition which may lead to negative kinetic energy and may even lead to a possibility of a negative total energy. This implies that negative energy particles can spontaneously come into existence and cause considerabe problems pertaining to stability and causality.
 
Again, on the other hand, if we consider the Viaggiu entropy given in Eq. (\ref{entropy-2}) and the Hawking temperature given in Eq. (\ref{htapp-2}) and note that $dU=0$, then
\begin{equation}
\dot{S}_{TA} = 6\pi R_A \frac{(1+w)(8+3w)}{(1-3w)},
\end{equation}
which suggests that the GSL is satisfied in one of the following four cases: (i) $-1 \leq w \leq \frac{1}{3}$, (ii) $w < -\frac{8}{3}$, (iii) $-\frac{8}{3} \leq w \leq -1$ and $w>\frac{1}{3}$, and (iv) $w \leq -\frac{8}{3}$ and $w>\frac{1}{3}$. Here, we observe that for dust, $p=0$, so $w=0$, and $\dot{S}_{TA} = 48 \pi R_A$, i.e., the GSL always holds. Now, none of the last three cases includes the dust scenario and hence they may be safely neglected. Therefore, in this scenario, GSL will be satisfied if $-1 \leq w \leq \frac{1}{3}$ which again means that the phantom era is forbidden. We are now in a position to tabulate (see Table I) the above results for a quick comparison of the thermodynamic implications of the Bekenstein and the Viaggiu entropies in a flat FLRW universe endowed with a perfect fluid having a constant EoS parameter $w$.\\\\
{\bf Table I}: Study of GSL with the Bekenstein and Viaggiu formalisms. The mathematical expressions in each case correpond to the respective first order time-derivative of the total entropy, i.e., $\dot{S}_{TA}$
\begin{center}
\begin{tabular}{|p{4cm}|p{6.5cm}|p{6.5cm}|}
\hline \hline & \begin{center} \textbf{Hawking temperature (Eq. (\ref{htapp-2}))} \end{center} & \begin{center} \textbf{Truncated Hawking temperature (Eq. (\ref{htapp-3}))} \end{center}\\
\hline \hline \begin{center} \textbf{Bekenstein entropy} \end{center} & \begin{center} $18\pi R_A \frac{(1+w)^2}{(1-3w)}$ \end{center} & \begin{center} $\frac{9\pi R_A}{2} (1+w)^2$ \end{center}\\
\hline \begin{center} \textbf{Remarks} \end{center} & GSL holds for $w \leq \frac{1}{3}$, that is, GSL holds in every epoch except during the pre-radiation era & GSL holds without any restriction on $w$\\
\hline \hline \begin{center} \textbf{Viaggiu entropy} \end{center} & \begin{center} $6\pi R_A \frac{(1+w)(8+3w)}{(1-3w)}$ \end{center} & \begin{center} $\frac{3\pi R_A}{2}(1+w)(8+3w)$ \end{center}\\
\hline \begin{center} \textbf{Remarks} \end{center} & GSL holds if $-1 \leq w \leq \frac{1}{3}$ and the phantom era is thermodynamically forbidden & GSL holds either if $w \geq -1$ or if $w \leq -\frac{8}{3}$ and the phantom era is thermodynamically forbidden\\
\hline \hline
\end{tabular}
\end{center}

\section{Short discussion}
This paper dealt with a study of the GSL by considering the dynamical apparent horizon to be endowed with the Hawking temperature and the Viaggiu entropy. The latter is a corrected form of the Bekenstein entropy obtained due to the dynamical degrees of freedom of the expanding Universe. We devoted our attention to a flat FLRW universe filled with a perfect fluid having a constant equation of state $p=w\rho$. We evaluated the first time-derivative of the total entropy of the Universe bounded by the dynamical apparent horizon assuming the Viaggiu formalism and determined the conditions which must be satisfied by the EoS parameter $w$ so that the GSL is satisfied. We then compared our results with those obtained with the Bekenstein formalism. It is worthwhile to note that we have considered both forms of the Hawking temperature, the original one as well as the truncated version. Our analysis yielded a startling result. {\it The Viaggiu entropy naturally forbids the phantom era. In other words, the Viaggiu entropy will never allow the EoS parameter to go beyond the Cosmological Constant.} This is in agreement with recent observations. Moreover, this result overcomes the problems of considering a phantom fluid. The essence of our work is that we have been able to deduce this result purely by thermodynamic means and is in sharp contrast with the results obtained with the Bekenstein entropy. However, following Viaggiu's work \cite{Viaggiu2}, we find that if either an open or a closed universe is considered, then the internal energy does not vanish at the dynamical apparent horizon. This means that our assumption that $dU=0$ will no longer be valid in such universes and it might lead to a whole new picture thermodynamically. This problem shall be dealt with in a future work.


\begin{acknowledgments}

The author is grateful to several anonymous reviewers for their valuable comments and suggestions which have helped to enchance the quality of the manuscript significantly.

\end{acknowledgments}


\end{document}